\documentclass[pre,aps,superscriptaddress,showpacs]{revtex4}

\def \beq{\begin{equation}}         \def \eeq{\end{equation}}
\def \beqa{\begin{eqnarray}}        \def \eeqa{\end{eqnarray}}
\def \bea{\begin{array}}        \def \eea{\end{array}}

\usepackage{amsmath}
\usepackage{epsfig}
\usepackage[dvips]{color}

\begin{document}

\title{A generalized integral fluctuation theorem for diffusion processes}
\author{Fei Liu}
\email[Email address:]{liufei@tsinghua.edu.cn} \affiliation{Center
for Advanced Study, Tsinghua University, Beijing, 100084, China}
\author{Zhong-can Ou-Yang}
\affiliation{Center for Advanced Study, Tsinghua University,
Beijing, 100084, China} \affiliation{Institute of Theoretical
Physics, The Chinese Academy of Sciences, P.O.Box 2735 Beijing
100080, China}
\date{\today}

\begin{abstract}
{We present a generalized integral fluctuation theorem (GIFT) for
general diffusion processes using the Feynman-Kac and
Cameron-Martin-Girsanov formulas. Existing IFTs can be thought of
to be its specific cases. We interpret the origin of this theorem
in terms of time-reversal of stochastic systems.}
\end{abstract}
 \maketitle

\section{Introduction}
Feynman-Kac (FK) formula, originally found by Feynman in quantum
mechanics~\cite{Feynman} and extended by Kac~\cite{Kac},
establishes an important connection between partial differential
equations (PDEs) and stochastic processes. Using this formula, one
may solve certain PDE by simulating a stochastic process.
Recently, the remarkable FK formula was employed by Hummer and
Szabo (HS)~\cite{Hummer01} to prove the celebrated Jarzynski
equality (JE)~\cite{JarzynskiPRE97,JarzynskiPRL97} in
nonequilibrium statistics~\cite{exp0}. Their work not only
provided a concise proof but also pointed out that intrinsic free
energy surfaces of biomolecules could be efficiently extracted by
nonequilibrium single-molecule manipulation experiments.

Compared to extensive interests of applications of HS's work in
single-molecule biophysics~\cite{Ritort,Bustamante}, little
attention was paid to their proving method. Very recently, Ge and
Jiang (GJ) reinvestigated previous derivation from a rigorously
mathematic point of view~\cite{Ge}. They pointed out that HS
misused the FK formula since it is usually applied for Kolmogorov
backward equation~\cite{Stroock} rather than Kolmogorov forward
(or Fokker-Planck) equation. In fact, Chetrite and Gawedzki
(CG)~\cite{Chetrite} have given correct proof of the JE using the
FK formula slightly earlier. Interestingly, however, these two
works seem to be apparently distinct though they employed the same
mathematic formula: GJ constructed a simple time-invariable
integral (the proof of theorem 2.6 in Ref.~\cite{Ge}), whereas CG
was based on time-reversal concept.

In addition to the JE, there are serval analogous equalities found
in past several
years~\cite{Crooks99,Crooks00,Maes,SeifertPRL05,HatanoSasa,Speck,Schmiedl}.
All of them have a type of
\begin{eqnarray}
\langle \exp[-{\cal A}]\rangle=1, \label{IFT}
\end{eqnarray}
where $\cal{A}$ is a functional of the stochastic trajectory of a
stochastic system, and angular brackets denote an average over the
ensemble of the trajectories. For instance, $\cal{A}$ is total
entropy change along a trajectory~\cite{Maes}. These equalities
were called integral fluctuation theorem (IFTs) to distinguish the
detailed fluctuation theorem
(DFTs)~\cite{Evans,Gallavotti,Kurchan,Lebowitz,Crooks00}. Due to
the similarity between the IFTs with the JE in formality, one may
naturally consider whether these equalities or even more general
versions could be derived by GJ's approach as well. On the other
hand, it should be intriguing to clarify the relationship between
their approach and the others, {\emph e.g.} time-reversal, which
was not performed in the previous work. In this work,
we present our results about these two questions.
With the help of the FK and Cameron-Martin-Girsanov
(CMG)~\cite{Cameron,Girsanov} formulas, we obtain a generalized
integral fluctuation theorem. We find it could be explained from
the concept of time-reversal.

\section{A time-invariable integral}
We consider a general $N$-dimension stochastic system ${\bf
x}=\{x_i\}, i=1,\cdots, N$ described by a stochastic differential
equation (SDE)~\cite{Gardiner}
\begin{eqnarray}
d{\bf x}(t)={\bf A}({\bf x},t)dt + {\bf B}^{\frac{1}{2}}({\bf
x},t)d{\bf W}(t) \label{SDE}
\end{eqnarray}
where $d{\bf W}$ is an $N$-dimensional Wiener process, ${\bf
A}=\{A_i\}$ denotes a $N$-dimensional drift vector and, ${\bf
B}^{1/2}$ is the square root of a $N$$\times$$N$ positive definite
diffusion matrix ${\bf B}=\{B_{ij}\}$. We use Ito's convention for
stochastic differentials. Rather than directly solving
Eq.~(\ref{SDE}), one usually converts the SDE into two equivalent
PDEs of the conditional probability $\rho({\bf x},t|{\bf x}',t')$
($t>t'$): the forward Fokker-Planck equation (FPE) $\partial_t
\rho= {\cal L}({\bf x},t)\rho$ with
\begin{eqnarray}
{\cal L}({\bf x},t)=-\partial_{x_i} A_i({\bf x},t)
+\frac{1}{2}\partial_{x_i}\partial_{x_j} B_{ij}({\bf x},t) ,
\label{FKoperator}
\end{eqnarray}
and the backward FPE $\partial_{t'} \rho = -{\cal L}^+({\bf
x}',t') \rho$ with
\begin{eqnarray}
{\cal L}^+({\bf x},t)=A_i({\bf x},t)\partial_{x_i} +
\frac{1}{2}B_{ij}({\bf x},t)\partial_{x_i}\partial_{x_j}.
\label{BFKoperator}
\end{eqnarray}
The initial conditions in both cases are $\rho({\bf x},t|{\bf
x}',t)=\delta({\bf x}-{\bf x}')$. The two operators ${\cal L}$ and
${\cal L}^{+}$ are adjoint each other. Here we use Einstein's
summation convention throughout this work unless explicitly
stated.

Based on the symbols and definitions above, we find that, if
$u({\bf x},t)$ satisfies a partial differential equation
\begin{eqnarray}
\label{PDF}
\partial_{t'} u({\bf x},t')&=&-{\cal L}^{+}({\bf x},t')u({\bf x},t')\\
&-&f^{-1}({\bf x},t')\left[ \partial_{t'} f({\bf x},t')-
{\cal L}({\bf x},t')f({\bf x},t')\right]u({\bf x},t') \nonumber\\
&+&f^{-1}({\bf x},t')\left[{\cal L}_a({\bf x},t')g({\bf
x},t')-g({\bf x},t'){\cal L}_a^{+}({\bf x},t')\right] u({\bf
x},t'),\nonumber
\end{eqnarray}
where $f({\bf x},t')$ and $g({\bf x},t')$ are arbitrary normalized
smooth positive functions~\cite{exp1}, ${\cal L}_a$ and ${\cal
L}_a^{+}$ are adjoint operators like Eqs.~(\ref{FKoperator}) and
(\ref{BFKoperator}), which may be the same or different with those
of the system we focus on, we have
\begin{eqnarray}
\frac{d}{dt'}\int d{\bf x}f({\bf x},t')u({\bf x},t')=0.
\label{GIdentity}
\end{eqnarray}
The proof is obvious if one makes use of the adjoint property of
these operators and notes that the derivative of the integral with
respect to time is
\begin{eqnarray}
\frac{d}{dt'}\int d{\bf x} f({\bf x},t')u({\bf x},t')=\int d{\bf
x}\left[ f\partial_{t'} u({\bf x},t')+u\partial_{t'} f({\bf
x},t')\right]. \label{demo}
\end{eqnarray}
This is simply generalization of GJ's idea, where the last three
terms in Eq.~(\ref{PDF}) were absent~\cite{exp2}. Multiplying both
sides of the PDF by $f({\bf x},t)$, one may see that this equation
has a certain symmetry with respect to the functions $f$ and $u$.
In the following, we investigate Eq.~(\ref{PDF}) with Liouvill-
and Fokker-Planck-type ${\cal L}_a$, respectively. Although one
may think that the former is only a specific case of the latter
(vanishing diffusion matrix), the Liouvill-type may be more
intriguing that will be seen shortly.

\section{Liouvill-type ${\cal L}_a({\bf x},t)$}
We assume
\begin{eqnarray}
{\cal L}_ag({\bf x},t')=2\partial_{x_i} S_i({\bf x},t'),
\end{eqnarray}
and ${\bf S}=\{S_i({\bf x},t)\}$ is an arbitrary vector having
natural boundary condition. The coefficient $2$ is for convenience
in discussion. We rewrite Eq.~(\ref{PDF})
\begin{eqnarray}
\label{LiouvillPDF}
\partial_{t'} u({\bf x},t')&=&-{\cal L}^{+}({\bf x},t')u({\bf x},t')\\
&-&f^{-1}({\bf x},t')\left[ \partial_{t'} f({\bf x},t')-
{\cal L}({\bf x},t')f({\bf x},t')\right]u({\bf x},t') \nonumber\\
&+&2f^{-1}({\bf x},t')\left[\partial_{x_i} S_i({\bf
x},t')+S_i({\bf x},t')\partial_{x_i}\right] u({\bf
x},t').\nonumber
\end{eqnarray}
Using the FK and CMG formulas~\cite{Stroock}, we obtain a
stochastic representation of the solution $u({\bf x},t)$ of the
PDF given by
\begin{eqnarray}
u({\bf x},t')=E^{{\bf x},t'}\left\{e^{-{\cal J}[f,{\bf S}, {\bf
x}(\cdot)] }q[{\bf x}(t)]\right\}\label{PDFsolution}.
\end{eqnarray}
where $q(\bf x)$ is the final condition of $u({\bf x},t)$, and the
functional
\begin{eqnarray}
\label{functional} {\cal J} [f,{\bf S}, {\bf x}(\cdot)]
&=&\int_{t'}^t\frac{1}{f}\left[\left({\cal
L}-\partial_\tau\right)f+2\partial_{x_i}S_i +\frac{2}{f} S_i({\bf
B}^{-1})_{ij} S_j\right]
d\tau\nonumber \\
&+&\int_{t'}^t \frac{2}{f}S_i\left({\bf
B}^{-1}\right)_{ij}\left(dx_j-A_id\tau\right),
\end{eqnarray}
where the expectation $E^{{\bf x},t'}$ is an average over all
trajectories ${\bf x}(\cdot)$ determined by Eq.~(\ref{SDE}) taken
conditioned on ${{\bf x}(t')={\bf x}}$, and the last term is Ito
stochastic integral. Combining Eqs.~(\ref{GIdentity})
and~(\ref{PDFsolution}), we have
\begin{eqnarray} \int d{\bf x'}
f({\bf x'},0)E^{{\bf x}',0}\left\{e^{-{\cal J}[f,{\bf S}, {\bf
x}(\cdot)] }q[{\bf x}(t)]\right\}=\int d{\bf x} f({\bf} x,t)q({\bf
x}),\label{GIFT}
\end{eqnarray}
by choosing $t'=0$. This equation has the same form as
Eq.~(\ref{IFT}) for $q({\bf x})=1$. We call it generalized
integral fluctuation theorem (GIFT) in this work. Note other
functions of $q({\bf x})$ is still useful in practice.

\subsection{Relationship between GIFT and existing IFTs}
Existing several IFTs are specific cases of Eq.~(\ref{GIFT}).
First case is to choose $S_i=0$ and $f=p^{\rm ss}({\bf x},t)$,
where $p^{ss}({\bf x},t')$ is transient steady-state solution
\begin{eqnarray}
{\cal L}({\bf x},t')p^{ss}({\bf x},t')=0
\end{eqnarray}
Then the functional~(\ref{functional}) is simply
\begin{eqnarray}
{\cal J}=-\int_0^t \partial_\tau \ln p^{\rm
ss}\left[x(\tau),\tau\right]d\tau.
\end{eqnarray}
GJ has analyzed this case with a specific final condition $q({\bf
x})=1$ in great details. According to the system driven by a
time-dependent conservative (a gradient of a potential) or
non-conservative force, Eq.~(\ref{GIFT}) reduces to Jarzynski
equality~\cite{JarzynskiPRE97,JarzynskiPRL97} and Hatano-Sasa
relation~\cite{HatanoSasa}, respectively~\cite{Ge}. Here we are
not ready to repeat the same derivatives but only to point out the
necessary of the general final condition: one obtains the key
Eq.~(4) in the HS's work only choosing $q({\bf x})=\delta({\bf
x}-{\bf z})$.

The second case is to choose $S_i$ to be the probability current
$J_i[f]$ of arbitrary normalized function $f({\bf x},t)$, where
\begin{eqnarray}
J_i[f({\bf x},t)]=A_i({\bf x},t)f({\bf
x},t)-\frac{1}{2}\partial_{x_j} B_{ij}f({\bf x},t).
\label{probcurrent}
\end{eqnarray}
Substituting the current into Eq.~(\ref{functional}) and doing
some derivations (details see the Appendix), we obtain
\begin{eqnarray}
{\cal J}= \ln \frac{f[{\bf x'(t')},t']}{f[{\bf x(t)},t]} + 2(\rm
S)\int_{t'}^t\hat{A}_i\left({\bf B^{-1}}\right)_{ij}
\left[\dot{x}_j -\frac{1}{2}({\bf
B}^{-\frac{1}{2}})_{ks}\partial_{x_k}({\bf B}^{-\frac{1}{2}})^{\rm
T}_{sj}\right] d\tau \label{totentropy}
\end{eqnarray}
where $\hat{A}_i=A_i-\partial_{x_l}B_{il}/2$ and
$\dot{x}_j=dx_j/d\tau$. Note that we used the Stratonovich'
integral (denoted by ``S" therein). Particularly, if the diffusion
matrix is constant that is usually assumed in Langevin equation
and $f({\bf x},t)$ is the solution of the stochastic system with a
initial condition $f({\bf x'},0)$, the functional is just the
total entropy change $\Delta s_{\rm tot}$: the first term in
Eq.~(\ref{totentropy}) is the system entropy change, and the
second is the entropy change of environment~\cite{SeifertPRL05}.
Hence, Eq.~(\ref{GIFT}) is the IFT of the total
entropy~\cite{SeifertPRL05,Maes} in this case.

The last case is for the system being a nonequilibrium
steady-state with a distribution $p^{\rm  ss}({\bf x})$. We choose
$f=p^{\rm ss}({\bf x})$ and $S_i=J_i(p^{\rm ss})$. Considering
that $\partial_{x_i} S_i=0$, the functional becomes the
housekeeping heat~\cite{Speck}
\begin{eqnarray}
{\cal J}= 2\int_{t'}^t\frac{J_i(p^{\rm ss})}{p^{\rm ss}}\left({\bf
B}^{-1}\right)_{ij}\left[\dot{x}_j -\frac{1}{2}({\bf
B}^{-\frac{1}{2}})_{kl}\partial_{x_k}({\bf B}^{-\frac{1}{2}})^{\rm
T}_{lj}\right] d\tau \label{housekeeping}
\end{eqnarray}
Of course, we can obtain this functional from
Eq.~(\ref{totentropy}) directly. Importantly, this IFT is even
correct for time-dependent case~\cite{Speck}. Although
Eq.~(\ref{GIFT}) does not include this equality, one can indeed
prove it using CMG formula~\cite{Chetrite}.

\section{Fokker-Planck-type ${\cal L}_a({\bf x},t')$}
If the diffusion matrix of ${\cal L}_a$ is not zero, we can still
obtain a GIFT. Given
\begin{eqnarray} {\cal L}_a({\bf x},t') =\left[-\partial_{x_i} a^i({\bf x},t') +
\frac{1}{2}\partial_{x_i}\partial_{x_j} b_{ij}({\bf x},t')\right].
\label{FPtype}
\end{eqnarray}
Assuming $a^i=a_1^i({\bf x},t)-2a_2^i({\bf x},t)$, we split this
operator into a sum of a Fokker-Planck- and Liouvill operators,
{\emph i.e.}
\begin{eqnarray}
{\cal L}_a({\bf x},t')=\left[{\cal L}^1_a({\bf
x},t')+2\partial_{x_i}a_2^i\right], \label{FPtypesplit}
\end{eqnarray}
where ${\cal L}^1_a$ is the same with Eq.~(\ref{FPtype}) except
that $a^i$ is replaced by $a_1^i$. Substituting
Eq.~(\ref{FPtypesplit}) into Eq.~(\ref{PDF}) and doing a simple
rearrangement, one may obtain Eq.~(\ref{LiouvillPDF}) again if we
redefine the drift vector and diffusion matrix of the operator
${\cal L}({\bf x},t')$ to be
\begin{eqnarray}
A'_i&=&A_i({\bf x},t')+a^1_i({\bf x},t')g({\bf x},t')/f({\bf
x},t')\nonumber \\
B'_{ij}&=&B_{ij}({\bf x},t')+b_{ij}({\bf x},t')g({\bf
x},t')/f({\bf x},t'),
\end{eqnarray}
and $S_i=a_2^i({\bf x},t')g({\bf x},t')$. Hence, we have a
GIFT~(\ref{GIFT}) except that the expectation $E^{x,t'}$ is now an
average over the trajectories generated from a new stochastic
system
\begin{eqnarray}
d{\bf x}(t)={\bf A'}({\bf x},t)dt + {\bf B'}^{\frac{1}{2}}({\bf
x},t)d{\bf W}(t). \label{NSDE}
\end{eqnarray}
From this aspect, Eq.~(\ref{GIdentity}) with the
Fokker-Planck-type ${\cal L}_a$ seems to not provide additional
information compared to the Liouvill-type~\cite{exp3}.

\section{GIFT and time reversal} We must emphasize that the
GIFT~(\ref{GIFT}) is derived mathematically (adjoint property of
the FK operators) without resorting to any physical reason. It
would be intriguing to establish a relationship between the
equality and time-reversal like that done by Crook~\cite{Crooks00}
in interpreting JE~\cite{JarzynskiPRE97,JarzynskiPRL97}. CG
recently suggested that the different fluctuation relations in
literature may be traced to different time
reversals~\cite{Chetrite}. In the remainder of the work, we use
the same viewpoint to understand the origin of Eq.~(\ref{PDF}) and
the GIFT.

We first give a definition of a time reversal~\cite{Chetrite}. The
variable $x_i$ of the stochastic system is even or odd according
to their rules under time reversal $t'\to t-t'$: if $x_i\to
y_i=+x_i$ is even and $x_i\to y_i=-x_i$ is odd. We write them in
abbreviation $x_i\to y_i=\varepsilon_i x_i$ and
$\varepsilon_i=\pm1$. The drift vector splits into ``irreversible"
and ``reversible" parts, ${\bf A}={\bf A}^{\rm irr}+{\bf A}^{\rm
rev}$. Under a time reversal, these vectors are transformed into
$\tilde{{\bf A}}=\tilde{{\bf A}}^{\rm irr}+{\tilde{\bf A}}^{\rm
rev}$, where
\begin{eqnarray} \tilde{A}_i^{\rm irr}({\bf x},
t')&=&\varepsilon_iA_i^{\rm
irr}({\bf y}, s), \\
\tilde{A}_i^{\rm rev}({\bf x}, t')&=&-\varepsilon_iA_i^{\rm
rev}({\bf y}, s),
\end{eqnarray}
${\bf y}=\{y_i\}$ and $s=t-t'$ ($0\leq t'\leq t$). We also define
a transformation of the diffusion matrix to be
\begin{eqnarray}
\tilde{B}_{ij}({\bf x},t')&=&\varepsilon_i\varepsilon_j
B_{ij}({\bf y}, s).
\end{eqnarray}
Note that no summation over repeated indices here. Because the
splitting is completely arbitrary, one may think it to be an
equivalent definition of a certain time reversal.

Considering a forward FPE $\partial_s p({\bf y},s)={\tilde{\cal
L}}({\bf y},s)p({\bf y},s)$ with the time-reversed drift vector
and diffusion matrix,
\begin{eqnarray}
{\tilde{\cal L}}({\bf y},s)=-\partial_{y_i} {\tilde A}_{i}({\bf
y},s)+\frac{1}{2}\partial_{y_i}\partial_{y_j}{\tilde B}_{ij}({\bf
y},s).\label{RFKoperator}
\end{eqnarray}
Substituting
\begin{eqnarray}
p({\bf y},s)=f({\bf x},t')v({\bf x},t')
\label{generaldetailbalance}
\end{eqnarray}
into above equation and doing some simple derivations, we obtain
\begin{eqnarray}
\label{SPDF}
\partial_{t'} v({\bf x},t')&=&-{\cal L}^{+}({\bf x},t')v({\bf x},t')\\
&-&f^{-1}({\bf x},t')\left[ \partial_{t'} f({\bf x},t')-
{\cal L}({\bf x},t')f({\bf x},t')\right]v({\bf x},t') \nonumber\\
&+&2f^{-1}({\bf x},t')\left[\partial_{x_i}S^{\rm irr}_i(f)+S^{\rm
irr}_i(f)\partial_{x_i}\right] v({\bf x},t'),\nonumber
\end{eqnarray}
where an irreversible probability current is defined to be
\begin{eqnarray}
S_i^{\rm irr}(f)= A^{\rm irr}_i({\bf x},t')f({\bf
x},t')-\frac{1}{2}
\partial_{x_k}B_{ik}({\bf x},t')f({\bf x},t')
\end{eqnarray}
We immediately see that Eq.~(\ref{LiouvillPDF}) would be the same
with Eq.~(\ref{SPDF}) if $S_i=S^{\rm irr}_i(f)$. Under this
circumstance, Eq.~(\ref{GIFT}) is a obvious consequence of the
specific time reversal of the stochastic system in that the
integral of the left-hand side of Eq.~(\ref{GIdentity}) equals to
$\int d{\bf y} p({\bf y},s)$, which is conservative with respect
to time. On the other hand, considering that the splitting of the
drift vector is completely arbitrary, we can define a time
reversal for a given vector $S_i$ by
\begin{eqnarray}
A^{\rm irr}({\bf x},t')&=&\frac{1}{f({\bf x},t')}
\left[S_i({\bf x},t')+\frac{1}{2}\partial_{x_k}B_{ik}({\bf x},t')f({\bf x},t') \right]\\
A^{\rm rev}{\bf x},t')&=&A_i({\bf x},t')-A^{\rm irr}({\bf x},t')
\end{eqnarray}
Then the GIFT for any vector ${\bf S}$ is a consequence of the new
defined time reversal of the stochastic system. The reader should
be reminded that these time reversals are not always meaningful or
easily realizable in physics. Before ending the section, we give
two comments about Eq.~(\ref{generaldetailbalance}). First, for a
homogenous diffusion process that satisfies the detailed balance,
this equation has been used earlier to connect the forward FK
solution $p({\bf x},t)$ and backward FK solution $v({\bf x},t)$,
where $f({\bf x},t)=p^{\rm eq}({\bf x})$~\cite{Gardiner}. Second,
Eqs.~(\ref{generaldetailbalance}) and the stochastic
representation solution $v({\bf x},t')$ [Eq.~(\ref{PDFsolution}]
are crucial for achieving DFTs~\cite{Crooks99,Crooks00}. CG has
given a detailed discussion about this point (Proposition 1
therein)~\cite{Chetrite}.

\section{Conclusion}
In this work, we present a generalized integral fluctuation
theorem for diffusion processes using the famous FK and CMG
formulas. Although one might think that this result~(\ref{GIFT})
is almost trivial from the point of view of time-reversal that we
explained here, the derivation by constructing a time-invariable
integral is novel. We close this work by pointing out several
directions that we can further improve the current results. First
is to give rigorous conditions for applying FK and CMG formulas.
Second, we may extend the results for the continuous diffusion
processes to discrete state case, {\emph e.g.} chemical master
equation. Finally, the most interesting to us is to apply the GIFT
in practical problems, {\emph e.g.} free energy reconstruction of
single-molecule manipulation.
\\

{\noindent We appreciate Dr. Chetrite for sending their intriguing
work~\cite{Chetrite} to us. The discussion about the connection
between the GIFT and time-reversal is mainly inspired by them.
This work was supported in part by Tsinghua Basic Research
Foundation and by the National Science Foundation of China under
Grant No. 10704045 and No. 10547002.}

\appendix
\section*{Appendix: derivation of the total entropy}
Substituting Eq.~(\ref{probcurrent}) into Eq.~(\ref{functional})
and separating the terms into $f$-dependent and -independent
parts, we obtain
\begin{eqnarray}
\label{functionalcurrent} {\cal
J}&=&\int_{t'}^t\left[\partial_{x_i}\hat{A}_i+2\hat{A}_i({\bf
B}^{-1})_{ij}\hat{A}_j\right]d\tau+2\hat{A}_i({\bf
B}^{-\frac{1}{2}})_{ij}dW_j \nonumber\\
&-&\int_{t'}^t\left[ \partial_\tau \ln f + A_i\partial_{x_i} \ln f
+ \frac{1}{2}B_{ij}\partial_{x_i}\partial_{x_j}\ln f \right ]d\tau
+\left(\partial_{x_i}\ln f\right)({\bf B}^{\frac{1}{2}})_{ij}dW_j
\end{eqnarray}
Employing the SDE~(\ref{SDE}), the second integral in above
equation becomes
\begin{eqnarray}
\int_{t'}^t\left[\partial_\tau \ln f
+\frac{1}{2}B_{ij}\partial_{x_i}\partial_{x_j}\ln
f\right]d\tau+\partial_{x_i}\ln fdx_i,
\end{eqnarray}
which is just the system entropy change $\int_{t'}^t d\ln
f=\ln\left[f(x(t),t)/f(x(t'),t')\right]$ according to the Ito
formula~\cite{Gardiner}. The entropy change of environment in
Eq.~(\ref{totentropy}) is derived from the first integral in
Eq.~(\ref{functionalcurrent}) by converting the Ito's stochastic
integral into the Stratonovich's:
\begin{eqnarray}
\int_{t'}^t2\hat{A}_i({\bf B}^{-\frac{1}{2}})_{ij}dW_j=(\rm
S)\int_{t'}^t2\hat{A}_i({\bf B}^{-\frac{1}{2}})_{ij}dW_j-
\frac{1}{2}({\bf B}^{\frac{1}{2}})_{jk}\partial_{x_k}\left[2({\bf
B}^{-\frac{1}{2}})^{\rm T}_{ij}\hat{A}_i\right]d\tau.
\end{eqnarray}
An alternative method is based on time-reversal technique.
Previous discussion shows that for $u({\bf x},t')$ satisfying
Eq.~(\ref{LiouvillPDF}) and $S_i=J_i(f)$, we can define a
time-reversal with ${\bf A}^{\rm irr}={\bf A}$ and ${\bf A}^{\rm
rev}=0$. Then $f({\bf x},t')u({\bf x},t')=p({\bf y},s)$ is a
solution of the FPE~(\ref{RFKoperator}). Because $f$ is arbitrary,
we may choose $f=1$ and $w({\bf x},t')=p({\bf y},s)$. Hence we
have $u({\bf x},t')=w({\bf x},t')\exp[-\ln f({\bf x},t')]$.
Obviously, the functional of the stochastic representation of
$w({\bf x},t')$ is the first integral of
Eq.~(\ref{functionalcurrent}). Like the proof of the GIFT, this
derivation shows the power of time-reversal technique again.


\begin{thebibliography}{99}
\bibitem{exp0}
Lebowitz and Spohn have used the Feynman-Kac and
Cameron-Martin-Girsanov formula earlier in proving fluctuation
theorem in their important work (p.350)~\cite{Lebowitz}.
\bibitem{exp1}
This nomalization requirement of these functions is inessential
for our discussion.
\bibitem{exp2}
We note that, if the last term of Eq.~(\ref{PDF}) was absent and
$-{\cal L}^+$ was replaced by ${\cal L}$, the equation would be
analogous to Eq.~(59) derived by P. Ao [Commun. Theor. Phys. {\bf
49}, 1073 (2008)]. But his result has the same flaw as
HS's~\cite{Ge}.
\bibitem{exp3}
It should be intriguing to investiage whether the solution of
Eq.~(\ref{PDF}) with Eq.~(\ref{FPtype}) has a stochastic
representation on the original stochastic system~(\ref{SDE}).
\end{thebibliography}
\end{document}